\documentstyle[12pt,aps,amsfonts,preprint,tighten,axodraw]{revtex}
\input epsf

\newcommand{\beq}{\begin{equation}}
\newcommand{\eeq}{\end{equation}}
\newcommand{\beqs}{\begin{eqnarray}}
\newcommand{\eeqs}{\end{eqnarray}}

\newcommand{\lsim}{\mathrel{\raisebox{-.6ex}{$\stackrel{\textstyle<}{\sim}$}}}
\newcommand{\gsim}{\mathrel{\raisebox{-.6ex}{$\stackrel{\textstyle>}{\sim}$}}}

\begin{document}
\draft
 
\baselineskip 6.0mm
 
\tighten
 
\title{Implications of Improved Upper Bounds on $|\Delta L|=2$ Processes}
 
\author{Laurence S. Littenberg \thanks{email: littenbe@bnl.gov}
\and Robert Shrock\thanks{email: robert.shrock@sunysb.edu, 
on sabbatical leave from Yang Institute for
Theoretical Physics, State University of New York, Stony Brook}}
 
\address{Physics Department \\
Brookhaven National Laboratory \\
Upton, NY  11973}
 
\maketitle
 
\vspace{10mm}
 
\begin{abstract}

We discuss implications of improved upper bounds on the $|\Delta L|=2$
processes (i) $K^+ \to \pi^- \mu^+ \mu^+$, from an experiment at BNL, and 
(ii) $\mu^- \to e^+$ conversion, from an experiment at PSI.  In particular, we
address the issue of constraints on neutrino masses and mixing, and on 
supersymmetric models with $R$-parity violation.

\end{abstract}
 
 
\vspace{16mm}
 
\newpage

At present there are increasingly strong indications for neutrino oscillations
and hence neutrino masses and lepton mixing from the solar neutrino deficiency
and atmospheric neutrino anomaly \cite{nurev}.  The existence of lepton mixing
means that lepton family number is not a good symmetry.  Majorana neutrino 
masses occur generically, and violate total lepton number $L$ by 
$|\Delta L|=2$ units. However, so far, in contrast to the data suggesting 
lepton mixing, experimental searches for the violation of total lepton number 
have only set limits.  Among these are searches for the $|\Delta L|=2$ 
processes (i) neutrinoless double
beta ($0\nu 2\beta$) decay of nuclei and (ii) $\mu^- \to e^+$ conversion in the
field of a nucleus.  A third class of $|\Delta L|=2$ processes includes the
decays $K^+ \to \pi^- \ell^+ \ell^{\prime +}$, where $\ell^+\ell^{\prime
+}=e^+e^+$, $\mu^+e^+$, or $\mu^+\mu^+$ \cite{early,pdec}. In a 
previous work we considered these decays and, from a retroactive 
data analysis, set the first upper limit on one of them, namely \cite{ls,cpv}
\beq
BR(K^+ \to \pi^- \mu^+ \mu^+) < 1.5 \times 10^{-4}  \quad (90 \% \ {\rm CL}). 
\label{lslimit} 
\eeq
In \cite{ls} we also noted that 
rare $K$ decay experiments at BNL could greatly improve this limit and 
proposed a search for $K^+ \to \pi^-\mu^+\mu^+$ \cite{e787}. 
Among these experiments was BNL E865, which was searching for, and has 
now set a stringent upper limit on, the decay $K^+ \to \pi^+\mu^+e^-$ 
\cite{zeller}.  This experiment has also recently obtained the 90 \% CL 
upper limit \cite{e865} 
\beq
BR(K^+ \to \pi^- \mu^+ \mu^+) < 3.0 \times 10^{-9} \ . 
\label{e865limit}
\eeq
In the present paper we discuss the implications of this limit.  
In the context of current bounds on neutrinoless double beta decay and 
$\mu^- \to e^+$ conversion, we shall also consider the implications of two 
other 90 \% CL limits on $|\Delta L|=2$ decays from E865 \cite{e865}:
\beq
BR(K^+ \to \pi^- e^+ e^+) < 6.4 \times 10^{-10} 
\label{kpee}
\eeq
and
\beq
BR(K^+ \to \pi^- \mu^+ e^+) < 5.0 \times 10^{-10} \ . 
\label{kpme}
\eeq

We first discuss physics sources for these decays, concentrating on massive 
Majorana neutrinos and $R$-parity-violating supersymmetric (SUSY) theories.  
In a modern theoretical context, one generally expects nonzero
neutrino mass terms, of both Dirac and Majorana type. 
Let us denote the left-handed flavor 
vector of SU(2) $\times$ U(1) doublet neutrinos as $\nu_L = 
(\nu_e,\nu_\mu,\nu_\tau)_L$ and the right-handed vector of 
electroweak-singlet neutrinos as $N_R = (N_1,..,N_{n_s})_R$.  
The Dirac and Majorana neutrino mass terms can then be written compactly as
 \beq 
-{\cal L}_m = {1 \over 2}(\bar\nu_L \
\overline{N^c}_L) \left( \begin{array}{cc} M_L & M_D \\ (M_D)^T & M_R
\end{array} \right )\left( \begin{array}{c} \nu^{c}_R \\ N_R \end{array}
\right ) + h.c.
\label{numass}
\eeq 
where $M_L$ is the $3 \times 3$ left-handed Majorana mass matrix, $M_R$ is
a $n_s \times n_s$ right-handed Majorana mass matrix, and $M_D$ is the 3-row by
$n_s$-column Dirac mass matrix.  In general, all of these are complex, and
$(M_L)^T = M_L \ , \quad (M_R)^T = M_R$.  The diagonalization of the matrix in
eq. (\ref{numass}) then yields $3+n_s$ mass eigenstates, which are generically
nondegenerate Majorana neutrinos (degeneracies in magnitudes of eigenvalues can
yield Dirac neutrinos).  Writing the charged current in terms of mass
eigenstates as $J_\lambda = \bar\ell_L \gamma_\lambda \nu_L$, one has, in
particular, $\nu_\mu =\sum_{j=1}^{3+n_s}U_{\mu j}\nu_j$.  The seesaw mechanism
naturally yields a set of three light masses for the three known neutrinos,
generically of order $m_\nu \sim m_D^2/M_R$, and $n_s$ very large masses
generically of order $m_R$, for the electroweak-singlet neutrinos, where $m_D
\sim M_{EW}$ and $m_R >> M_{EW}$ denote typical elements of the matrices $M_D$
and $M_R$, and $M_{EW} \simeq 250$ GeV is the electroweak symmetry breaking
scale.  However, if one tries to fit all current neutrino experiments,
including not only the solar and atmospheric, but also the LSND data, then it
is necessary to include electroweak-singlet (``sterile'') neutrinos with masses
$<< M_R$ in order to achieve an acceptable fit.  In this case, the weak
eigenstate $\nu_\mu$ may contain significant components of mass eigenstates
beyond the usual three light ones $\nu_j$, $j=1,2,3$, and, it is {\it a priori}
possible that some of these mass eigenstates might have masses lying in the
theoretically ``disfavored'' intermediate range $m_D/m_R^2 << m_{\nu_j} << m_R$
(subject to both particle physics and astrophysical/cosmological constraints).
Whether or not such intermediate-mass neutrinos exist is ultimately an
empirical question that must be settled by experiment.  It is therefore of
continuing interest to address constraints from data on neutrino masses in this
intermediate mass region.  Since current indications from solar and atmospheric
data suggest neutrino masses in the seesaw-favored region and since,
independent of this, neutrino masses in the range of a few to several hundred
MeV can suppress large-scale structure formation and hence may be disfavored 
by cosmological constraints \cite{raffelt}, we concentrate mainly on neutrino 
masses smaller or larger than this range here. 

There are two types of lowest-order graphs involving massive neutrinos that 
contribute to the decay $K^+ \to \pi^- \ell^+\ell^{\prime +}$, where
$\ell^+ \ell^{\prime +}=\mu^+ \mu^+$, $\mu^+ e^+$, or $e^+e^+$, as shown in 
Fig. 1.

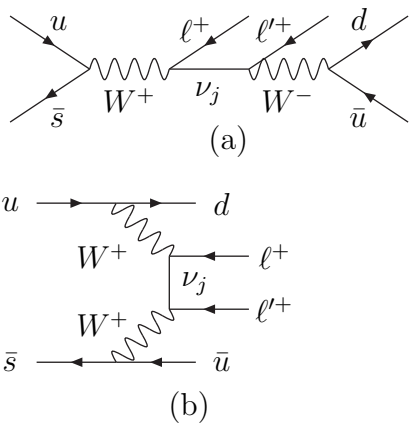
\begin{picture}(160,100)(0,0)
\ArrowLine(40,30)(10,10)
\ArrowLine(10,50)(40,30)
\Photon(40,30)(70,30){4}{4}
\Line(70,30)(100,30)
\ArrowLine(100,50)(70,30)
\ArrowLine(130,50)(100,30)
\Photon(100,30)(130,30){-4}{4}
\ArrowLine(130,30)(160,50)
\ArrowLine(160,10)(130,30)
\Text(25,45)[lb]{$u$}
\Text(25,15)[lt]{$\bar s$}
\Text(55,23)[t]{$W^+$}
\Text(80,45)[]{$\ell^+$}
\Text(85,25)[t]{$\nu_j$}
\Text(115,23)[t]{$W^-$}
\Text(110,45)[]{$\ell^{\prime +}$}
\Text(145,45)[rb]{$d$}
\Text(145,15)[rt]{$\bar u$}
\put(85,0){(a)}
\end{picture}

\begin{picture}(160,100)(0,0)
\ArrowLine(20,80)(50,80)
\ArrowLine(50,80)(80,80)
\ArrowLine(80,20)(50,20)
\ArrowLine(50,20)(20,20)
\Photon(50,20)(70,40){4}{4}
\Photon(50,80)(70,60){-4}{4}
\Line(70,40)(70,60)
\ArrowLine(100,60)(70,60)
\ArrowLine(100,40)(70,40)
\Text(10,20)[]{$\bar s$}
\Text(90,20)[]{$\bar u$}
\Text(10,80)[]{$u$}
\Text(90,80)[]{$d$}
\Text(80,50)[]{$\nu_j$}
\Text(110,40)[]{$\ell^{\prime +}$}
\Text(110,60)[]{$\ell^+$}
\Text(55,30)[rb]{$W^+$}
\Text(55,65)[rt]{$W^+$}
\put(70,0){(b)}
\end{picture}

\begin{figure}[h]
\caption{\footnotesize{Graphs involving massive Majorana neutrinos that
contribute to $K^+ \to \pi^- \ell^+ \ell^{\prime +}$, where
$\ell^+,\ell^{\prime +}=\mu^+\mu^+$, $\mu^+e^+$, or $e^+e^+$. 
In the case of identical $\ell^+$ and $\ell^{\prime +}$, it is understood 
that the contributions are from the diagrams minus the same diagrams with the 
outgoing antilepton lines crossed.}}
\label{fig1}
\end{figure}

\vspace{6mm}

For the $K^+ \to \pi^- \mu^+ \mu^+$ decay, the $s$-channel diagrams yield, in 
standard notation,
\beq
Amp(2(a)) = 2G_{F}^2 f_{K} f_{\pi} (V_{ud}V_{us})^* 
\sum_j (U_{\mu j}U_{\mu j})^* p_{_{K,\alpha}} p_{_{\pi, \beta}} 
\Bigl [ L_{j}^{\alpha \beta}(p_\mu, p_{\mu^\prime}) - 
L_{j}^{\alpha \beta}(p_{\mu^\prime}, p_{\mu}) \Bigr ] 
\label{amp1}
\eeq
where 
\beq
L_{j}^{\alpha \beta}(p_{\mu}, p_{\mu^\prime}) = 
m_{\nu_j}[q^2 - m_{\nu_j}^2]^{-1}
\bar v(p_{\mu})\gamma^{\alpha}\gamma^{\beta}P_R v^{c}(p_{\mu^\prime})
\label{L}
\eeq
where $q$ denotes the momentum carried by the virtual $\nu_j$, and 
$P_R=(1+\gamma_5)/2$ is a right-handed chiral projection operator. 
The $t$-channel graphs cannot be evaluated so easily, because the hadronic 
matrix element that occurs, 
\beq
\int d^4x d^4y e^{i(p_{d}-p_u) \cdot y} e^{i(p_{\bar s} - p_{\bar u}) \cdot x}
\langle \pi^- |[\bar d_{_L}(y) \gamma_{\beta} u_{_L}(y)] 
          [\bar s_{_L}(x) \gamma_{\alpha} u_{_L}(x)] |K^+ \rangle
\label{amp2}
\eeq
cannot be directly expressed in terms of measured quantities, unlike the 
matrix elements  $ \langle 0|\bar s_{_L} \gamma_{\alpha} u_{_L} |K^+ \rangle$ 
and $\langle \pi^-| \bar d_{L} \gamma_{\beta} u_{_L} |0 \rangle$ from the 
first graph. In the limits where $m_{\nu_j}^2$ 
is much smaller or larger than the magnitude of the typical $q^2$, 
$|\langle q^2 \rangle_{ave}| \sim O((10^2 \ {\rm MeV})^2)$,  the propagator 
factor simplifies:
\beq
\frac{m_{\nu_j}}{q^2 - m_{\nu_j}^2} \simeq \cases{ m_{\nu_j}/\langle q^2 
\rangle_{ave} & if $m_{\nu_j} << |\langle q^2 \rangle_{ave}|$ \cr
                    &   \cr
-1/m_{\nu_j} & if $m_{\nu_j} >> |\langle q^2 \rangle_{ave}|$ } \ . 
\label{propfac}
\eeq
Since the contribution of each virtual $\nu_j$ is accompanied by a complex
factor $(U_{\mu j}^*)^2$, it is possible for these contributions to add
constructively or destructively.
Because of the possibility of such cancellations, one cannot put an upper limit
on neutrino masses or lepton mixing matrix coefficients from an upper bound 
on the decay $K^+ \to \pi^- \mu^+ \mu^+$.  (A similar remark applies to 
$K^+ \to \pi^- \mu^+ e^+$ and $K^+ \to \pi^- e^+e^+$ since cancellations can
also occur among contributions of various $\nu_j$'s to the respective 
amplitudes for those decays.)  The same comment is well-known in the case of
neutrinoless double beta decay; for example, in the light-mass region, the
lower limit on half-lives for $0\nu2 \beta$ transitions places an upper limit
on $\sum_j U_{ej}^2 m_{\nu_j}$, not on $|U_{ej}|$ or $m_{\nu_j}$ themselves.

With these inputs, we estimated \cite{ls}
\beq
BR(K^+ \to \pi^- \mu^+ \mu^+) \sim 10^{-(13 \pm 2)}r_{\mu \mu}
\Bigl | \sum_j U_{\mu j}^2 f(m_{\nu_j}/(100 \ {\rm MeV}) \Bigr |^2 
\label{brestimate}
\eeq
where
\beq
f(z) =  \cases{ z & if $z << 1$ \cr
              1/z & if $z >> 1$ }
\label{fz}
\eeq
and the factor $r_{\mu \mu} \simeq 0.2$ is a relative phase space factor
(normalized relative to the decay $K^+ \to \pi^- e^+ e^+$).  If $z \sim 
O(1)$, one must, of course, retain the exact propagator.  In Fig. 1(a), the
range of (timelike) $q^2$ is $(m_{\pi^+}+m_\mu)^2 \le q^2 \le
(m_{K^+}-m_\mu)^2$, i.e., $245 \le \sqrt{q^2} \le 388$ MeV.  Hence, if there 
exists a neutrino with $m_{\nu_j}$ in this range, $245 \le m_{\nu_j} \le 
388$ MeV, then $q^2-m_{\nu_j}^2$ can vanish,
leading to a resonant enhancement of the amplitude for this graph \cite{dib}. 

In the following, we assume for simplicity that a single mass eigenstate 
$\nu_j$ dominates the sum in (\ref{brestimate}) but recall our remark above
concerning the possibility of cancellations and note that it is 
straightforward to generalize our
discussion to the case where there are several comparable contributions. 
If $m_{\nu_j} << m_K$, then, 
using the new upper bound (\ref{e865limit}) and the most
conservative choice for the numerical prefactor in eq. (\ref{brestimate}), 
(i.e., taking $10^{-(13 \pm 2)} \to 10^{-15}$), we obtain 
\beq
|U_{\mu j}|^2 \Bigl ( \frac{m_{\nu_j}}{100 \ {\rm MeV}} \Bigr )
< 4 \times 10^2 \biggl [ \frac{BR(K^+ \to \pi^-\mu^+\mu^+)}{3 \times 
10^{-9}} \Bigr ]^{1/2} \ . 
\label{newlimit1}
\eeq
On the other hand, if $m_{\nu_j} >> m_K$, we obtain 
\beq
|U_{\mu j}|^2 \Bigl ( \frac{100 \ {\rm MeV}}{m_{\nu_j}} \Bigr ) 
< 4 \times 10^2 \biggl [ \frac{BR(K^+ \to \pi^-\mu^+\mu^+)}{3 \times
10^{-9}} \Bigr ]^{1/2} \ .
\label{newlimit2}
\eeq
Since $m_{\nu_j} << m_K$ in order for (\ref{newlimit1}) to
hold, and since $|U_{\mu j}|<1$ by unitarity, the bound (\ref{newlimit1}) does 
not place a significant restriction on either of these quantities.  Similarly,
in the heavy neutrino mass region, since $m_{\nu_j} >> m_K$ in order for the 
bound (\ref{newlimit2}) to hold, it does not place any restriction on
$m_{\nu_j}$ or $|U_{\mu j}|$ in this mass region \cite{zuber}.  This is not to
say, however, that it is not worthwhile to search further for the decay 
$K^+ \to \pi^- \mu^+ \mu^+$, since it constitutes a testing ground for
violation of total lepton number that is quite different from the usual 
searches for neutrinoless double beta decay of nuclei. 

We also give an update of the limit on the decay $K^+ \to \pi^- \mu^+e^+$. In
\cite{ls} we obtained an indirect upper limit on this decay by
observing that the leptonic part of the amplitude for this decay is related by
crossing to the leptonic part of the amplitude for the conversion process in
the field of a nucleus $(Z,A)$: $\mu^- + (Z,A) \to e^+ + (Z-2,A)$.  At the time
of \cite{ls}, the best upper limit on this conversion process was 
$\sigma(\mu^- + Ti \to e^+ + Ca)/\sigma(\mu^- + Ti
\to \nu_\mu + Sc) < 1.7 \times 10^{-10}$ from a TRIUMF experiment
\cite{triumfmueconv}, and we inferred that 
$BR(K^+ \to \pi^- \mu^+e^+) \lsim \ {\rm few} \ \times 10^{-9}$.  
The current best bound, from a PSI experiment, is \cite{psimueconv}
\beq 
\frac{\sigma(\mu^- + Ti \to e^+ + Ca)}{\sigma(\mu^- + Ti \to \nu_\mu + Sc)} 
< 1.7 \times 10^{-12} \quad (90 \ \% \ {\rm CL}) \ . 
\label{mueconv}
\eeq
Using this new bound, we conservatively infer that 
\beq
BR(K^+ \to \pi^- \mu^+e^+) \lsim {\rm few} \ \times 10^{-11} \ . 
\label{kpmelimit}
\eeq
As was noted in \cite{ls}, this is an indirect limit since it requires a
theoretical estimate of the hadronic matrix element as input.  Our indirect
bound (\ref{kpmelimit}) is more stringent than the E865 limit (\ref{kpme}), 
but the latter limit is still valuable since it is direct.

One can obtain an indirect upper limit on the decay $K^+ \to \pi^-
e^+e^+$ from the existing upper limit on neutrinoless double beta
decay, which, for light neutrinos, gives $|\sum_j U_{ei}^2 m_{\nu_j}|
\lsim 0.4$ eV (depending on the input used for the nuclear matrix
elements) \cite{baudis}.  Using the same method as above, we obtain an
upper limit many orders of magnitude less than the direct limit
(\ref{kpee}).

One can also consider $\Delta L=2$ decays of $D$ and $B$ mesons.  Here
only rather modest direct upper limits of order $10^{-3}$ to $10^{-4}$
have been set on the branching ratios \cite{pdg}.  A similar comment
applies to $|\Delta L|=2$ hyperon decays, on which we previously set
upper limits \cite{hyp}.

We next proceed to discuss constraints on $R$-parity violating ($RPV$) 
SUSY models. 
$R$-parity may be defined as $R=(-1)^{3B+L+2S}$, where $B$, $L$, and $S$ refer
to the baryon and lepton numbers and to the spin of the particle \cite{susy}.
Although 
$R$-parity was originally hypothesized in order to prevent intolerably rapid
proton decay in SUSY models, one can achieve the same end 
by imposing weaker global
symmetries that forbid terms of the form $U^c_i D^c_j D^c_k$ 
in the superpotential (where the subscripts $i,j,k$ here are generation 
indices and we follow the usual convention of writing the holomorphic 
operator products in terms of left-handed chiral superfields) 
while still allowing the $R$-parity violating terms
\beq
W_{RPV}=\lambda_{ijk}L_i L_j E^c_k + 
\lambda^{\prime}_{ijk}L_i Q_j D^c_k + \kappa_i L_i H_u \ . 
\label{wrpv}
\eeq
These terms violate total lepton number and, in general, also lepton family
number.  A recent review of $R$-parity violating SUSY models is 
 \cite{dreiner}. 
The second term in (\ref{wrpv}) yields several contributions to $K^+ \to \pi^-
\mu^+\mu^+$, shown in Figs. 2(a,b), where $\tilde{\mu}$ and 
$\tilde\chi^0$ denote the scalar muon and neutralino.

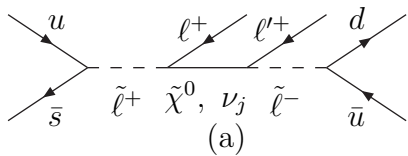
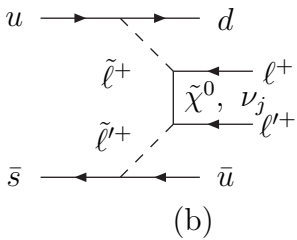
\begin{picture}(160,100)(0,0)
\ArrowLine(40,30)(10,10)
\ArrowLine(10,50)(40,30)
\DashLine(40,30)(70,30){4}
\Line(70,30)(100,30)
\ArrowLine(100,50)(70,30)
\ArrowLine(130,50)(100,30)
\DashLine(100,30)(130,30){4}
\ArrowLine(130,30)(160,50)
\ArrowLine(160,10)(130,30)
\Text(25,45)[lb]{$u$}
\Text(25,15)[lt]{$\bar s$}
\Text(55,23)[t]{$\tilde{\ell}^+$}
\Text(80,45)[]{$\ell^+$}
\Text(85,25)[t]{$\tilde{\chi}^0, \ \nu_j$}
\Text(115,23)[t]{$\tilde{\ell}^-$}
\Text(110,45)[]{$\ell^{\prime +}$}
\Text(145,45)[rb]{$d$}
\Text(145,15)[rt]{$\bar u$}
\put(85,0){(a)}
\end{picture}

\vspace{8mm}

\begin{picture}(160,100)(0,0)
\ArrowLine(20,80)(50,80)
\ArrowLine(50,80)(80,80)
\ArrowLine(80,20)(50,20)
\ArrowLine(50,20)(20,20)
\DashLine(50,20)(70,40){4}
\DashLine(50,80)(70,60){4}
\Line(70,40)(70,60)
\ArrowLine(100,60)(70,60)
\ArrowLine(100,40)(70,40)
\Text(10,20)[]{$\bar s$}
\Text(90,20)[]{$\bar u$}
\Text(10,80)[]{$u$}
\Text(90,80)[]{$d$}
\Text(90,50)[]{$\tilde{\chi}^0, \ \nu_j$}
\Text(110,40)[]{$\ell^{\prime +}$}
\Text(110,60)[]{$\ell^+$}
\Text(55,30)[rb]{$\tilde{\ell}^{\prime +}$}
\Text(55,65)[rt]{$\tilde{\ell}^+$}
\put(70,0){(b)}
\end{picture}

\vspace{8mm}

\begin{picture}(160,100)(0,0)
\ArrowLine(20,80)(50,80)
\ArrowLine(50,80)(80,80)
\ArrowLine(80,20)(50,20)
\ArrowLine(50,20)(20,20)
\DashLine(50,20)(70,40){4}
\DashLine(50,80)(70,60){4}
\Line(70,40)(70,60)
\ArrowLine(70,60)(100,60)
\ArrowLine(100,40)(70,40)
\Text(10,20)[]{$\bar s$}
\Text(90,20)[]{$\ell^{\prime +}$}
\Text(10,80)[]{$u$}
\Text(90,80)[]{$\ell^+$}
\Text(90,50)[]{$\tilde{\chi}^0 \ , \tilde{g}$}
\Text(110,40)[]{$\bar u$}
\Text(110,60)[]{$d$}
\Text(55,30)[rb]{$\tilde u^c$}
\Text(55,65)[rt]{$\tilde d$}
\put(70,0){(c)}
\end{picture}

\begin{figure}[h]
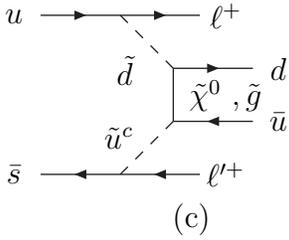

\caption{\footnotesize{Graphs that 
contribute to $K^+ \to \pi^- \ell^+ \ell^{\prime +}$, where
$\ell^+,\ell^{\prime +}=\mu^+\mu^+$, $\mu^+e^+$, or $e^+e^+$, in 
supersymmetric theories with $R_p$ violation.
In the case of identical $\ell^+$ and $\ell^{\prime +}$, it is understood
that the contributions are from the diagrams minus the same diagrams with the
outgoing antilepton lines crossed.}} 
\label{fig2}
\end{figure}

Note that there may be 
neutrino-neutralino mixing in $RPV$ theories, even if at some mass scale one 
rotates the terms $\kappa_i L_i H_u$ to zero.  A third type of diagram is shown
in Fig. 2(c).  As noted, for $K^+ \to \pi^- \mu^+ \mu^+$, each diagram is 
accompanied by minus the same diagram with the outgoing $\mu^+$ lines 
crossed.  The $\tilde \ell \tilde \chi^0 \mu$ vertices 
are $\propto \sqrt{g^2+g^{\prime 2}}$, where $g$ and $g^\prime$ are the SU(2)
and U(1)$_Y$ gauge couplings, while the $u d \tilde \mu$ and 
$\bar s \bar u \tilde \mu$ vertices are $\propto \lambda^{\prime}_{211}$ and 
$\lambda^{\prime}_{212}$, respectively and the $u d \tilde e$ and 
$\bar s \bar u \tilde e$ vertices are $\propto \lambda^{\prime}_{111}$ and
$\lambda^{\prime}_{211}$, respectively.  In the third type of diagram, 
there can be a gluino on the internal line, with $\tilde d \tilde g d$
and $\tilde u^c \tilde g \bar u$ vertices proportional to the strong coupling
$g_s$; or there can be a neutralino on the internal line, with vertices as
given above.  Bounds on $RPV$ couplings are model-dependent, but typical 
current upper bounds on $\lambda^\prime_{211}$, $\lambda^\prime_{212}$, 
$\lambda^\prime_{111}$, and $\lambda^\prime_{211}$ are $\lsim O(0.1)$ 
\cite{dreiner}.  Using these inputs, we find that these $R$-parity 
violating contributions could be much larger than those from massive neutrinos
and lepton mixing, but are still expected to be small compared with the 
upper limit (\ref{e865limit}):
\beq
BR(K^+ \to \pi^-\mu^+\mu^+)_{RPV} \lsim 10^{-16}
(\lambda^{\prime}_{211}\lambda^{\prime}_{212})^2 
\Bigl ( \frac{200 \ {\rm GeV}}{m_{SUSY}} \Bigr )^{10} \ . 
\label{rpvcontrib}
\eeq 
For the purpose of this rough estimate, we have taken the masses of the
various superpartners $\tilde u$, $\tilde d$, $\tilde \mu$, $\tilde \chi^0$,
and $\tilde g$ to be comparable and denoted this mass scale as $m_{SUSY} \sim
M_{EW}$.  Note that, given the lower bounds on the masses of $\tilde e$ and 
$\tilde \mu$ or order 100 GeV, no resonance is possible in the amplitude of 
Fig. 2(a). One could, of course, take a particular SUSY parameter set and
perform the calculation for this set; however, there is a large range of
variation in possible superpartner masses as well as allowed values of other
relevant parameters such as $\tan \beta$ (as illustrated, e.g., by the
parameter sets used in \cite{atlas}), so we deliberately keep our estimate
general.  Hence, it appears that the limit (\ref{e865limit}) does not strongly
constrain possible $RPV$ SUSY theories.  The constraints on $R$-parity
violating models from $\mu^- \to e^+$ conversion and neutrinoless double beta
decay are also more stringent than those from the limits (\ref{kpme}) and
(\ref{kpee}).

Thus, while neutrinoless double beta decay and $\mu^- \to e^+$ conversion are
the most sensitive ways to search for $|\Delta L|=2$ transitions with $|\Delta
L_e|=2$ and $\Delta L_e = \Delta L_\mu = \pm 1$, respectively, the decay $K^+
\to \pi^-\mu^+\mu^+$ is, at present, the best way to search for $|\Delta L|=2$
transitions with $|\Delta L_\mu|=2$. 
It is therefore worthwhile to estimate the potential of
future $K^+$ decay experiments to probe for this decay to lower values
of branching ratio.  In particular this might be undertaken by the CKM
experiment planned at Fermilab \cite{ckm}. The proposal for this
experiment anticipates a statistical sensitivity of $\sim
10^{-12}$/event for the decay $K^+ \to \pi^+ \nu\bar\nu$.  While it
would require a substantial, and as yet unplanned, effort to design
and build a trigger to search for $K^+ \to \pi^-\mu^+\mu^+$,
it appears \cite{ckm} that this experiment might be able to
reach a level near to $10^{-12}$ in branching ratio and thus improve
substantially on the already impressive upper limit on $BR(K^+ \to \pi^- \mu^+
\mu^+)$ from E865 at Brookhaven.  We also suggest that the planned experiment
MECO at BNL \cite{meco}, which anticipates searching for $\mu^- \to e^-$
conversion below the level $\sim 10^{-16}$ relative to $\mu^-$ capture,
should also undertake a search for $\mu^- \to e^+$ conversion.

\vspace{8mm}
 
    We thank H. Ma for discussions. The research of L.S.L. was supported by DOE
contract DE-AC02-98CH10886.  The research of R.E.S. was supported at BNL by the
DOE contract DE-AC02-98CH10886 and at Stony Brook by the NSF grant
PHY-97-22101. The U.S. government retains a non-exclusive royalty-free license
to publish or reproduce the published form of this contribution or to allow
others to do so for U.S. government purposes.

\vfill
\eject
 

\begin{thebibliography}{99}
\bibitem{pdec}{We focus here on $|\Delta L|=2$ processes; $|\Delta L|=1$ 
processes are also of interest, e.g., in the context of searches for nucleon
decay modes such as $p \to e^+ \pi^0$, $\mu^+ \bar K^0_s$, etc.}
\bibitem{cpv}{In Ref. \cite{ls} and here we neglect very small $CP$
violating effects that could make $\Gamma(K^+ \to \pi^- \ell^+
\ell^{\prime +})$ differ slightly from $\Gamma(K^- \to \pi^+ \ell^-
\ell^{\prime -})$.}
\bibitem{e787}{The BNL-Princeton-TRIUMF experiment E787, which has now observed
the rare decay $K^+ \to \pi^+ \nu\bar\nu$ (E787 Collaboration (S. Adler et
al.), Phys. Rev. Lett. {\bf 79} (1997) 2204; {\bf 84} (2000) 3768 and measured
$BR(K^+ \to \pi^+ \nu\bar\nu)=1.5^{+3.4}_{-1.2} \times 10^{-10}$, also has some
sensitivity for searching for $K^+ \to \pi^-\mu^+\mu^+$.  However, an estimate
of the upper limit that can be derived from E787 data is less stringent than
the limit (\ref{e865limit}) that has now been obtained by E865.}
\bibitem{zuber}{Our results here differ with the recent claim in
K. Zuber, Phys Lett {\bf B479} (2000) 33 (hep-ph/0003160). In Fig. 2
of that paper, an excluded region is shown in the region $m_{\nu_j}
\gsim 500$ GeV with the property that the larger the value of
$m_{\nu_j}$, the more stringent the upper bound on $|U_{\mu j}|^2$; in
fact, the opposite is the case: because of the propagator dependence
(\ref{propfac}), in this mass region, the larger the value of
$m_{\nu_j}$, the less stringent is the constraint on $|U_{\mu j}|^2$.
Moreover, as our analysis shows, there is is actually no constraint on
$|U_{\mu j}|^2$.  The erroneous claims in \cite{zuber} can be traced
to the author's use of the light-mass reduction of the propagator in
the heavy mass region where it does not apply.}
\end{thebibliography}
\end{document}